# Forecasting the Turkish Lira Exchange Rates through Univariate Techniques: Can the Simple Models Outperform the Sophisticated Ones?[1]

Mostafa R. Sarkandiz

*Graduate School of Applied Mathematics, Middle East Technical University, Turkey*
Mostafa.sarkandiz@metu.edu.tr

**Abstract**

Throughout the past year, Turkey's central bank policy to decrease the nominal interest rate has caused episodes of severe fluctuations in Turkish lira exchange rates. According to these conditions, the daily return of the USD/TRY have attracted the risk-taker investors' attention. Therefore, the uncertainty about the rates has pushed algorithmic traders toward finding the best forecasting model. While there is a growing tendency to employ sophisticated models to forecast financial time series, in most cases, simple models can provide more precise forecasts. To examine that claim, present study has utilized several models to predict daily exchange rates for a short horizon. Interestingly, the simple exponential smoothing model outperformed all other alternatives. Besides, in contrast to the initial inferences, the time series neither had structural break nor exhibited signs of the ARCH and leverage effects. Despite that behavior, there was undeniable evidence of a long-memory trend. That means the series tends to keep a movement, at least for a short period. Finally, the study concluded the simple models provide better forecasts for exchange rates than the complicated approaches.

## I) Introduction

In most macroeconomic analyses, the exchange rate has always been an integral part of the macro models because the rate plays a crucial role in determining the export/import ratio, which is one of the fundamental parameters in GDP formation and inflation fluctuations [1]. If there are no transaction costs and trade barriers, the purchasing power parity (PPP) hypothesis states the exchange rate of two currencies equals the ratio of their inflation rates. The hypothesis has been subjected to numerous investigations; however, it has been rejected in most cases. In fact, a PPP-based exchange rate time series that is more stable than the market series can be calculated, and in most cases, two series are cointegrated. For instance, [2] found out there is no short-term co-movement between those time series, but in the long run, the market rates tend to move toward the PPP rates. Actually, the short-term decoupling happens because more parameters other than inflation rate influence the currency's ratio. In this regard, [3], by conducting several diagnostic tests, concluded there is a negative long-term nexus between the balance of trade and exchange rate. In contrast, [4] found a strong positive relationship between them; however, in the short run, the correlation could be insignificant or non-linear. In addition to those mentioned factors, the foreign debts and the credit risk are two examples of other

---

influential variables [5]. As a result, since there is no universal theorem explaining the short-time relationship between exchange rates and macroeconomic parameters, employing univariate forecasting models would be reasonable, specifically if the daily time series is under investigation. For rationalization, according to the above survey, the inflation rate is one of the most influential parameters affecting the exchange rate path; however, there is no daily data for inflation rates. The only quasi-proxy for the daily rate is the return rate of breakeven inflation which is the return of the differences between a 10-year and a 10-year inflation-indexed Treasury bond yields. But this indicator can be considered as the inflation expectations rather than the actual rate [6]. However, since Turkey's government has not issued inflation-indexed bonds so far, employing this proxy is impractical. As a result, a multivariate model in the best-case scenario can provide monthly forecasts, which is not desired for daily transactions. Therefore, univariate models are the only feasible alternative.

Throughout the past year, the Turkish lira exchange rate has experienced massive downward trend. Some parts of the problem definitely can be attributed to the Covid-19 virus pandemic and a drastic fall in foreign income due to the extended lockdowns and almost the ban on entering foreign tourists. On the other hand, in response to the increasing rates of inflation, most countries have incremented the nominal interest rates as a tightening monetary policy to battle the higher levels of inflation. For example, the Federal Reserve Bank of the U.S. has gently increased the federal funds rate from 0.25 to 4.0 percent during the past nine months (Federal Reserve Bank of St. Louis[2]). In contrast, the central bank of Turkey during that period has declined the rate several times, especially from 11.5 in August to 7.50 in November of 2022 [3]. In conventional monetary theories, a decrease in the interest rate is considered an expanding policy that causes inflation rates to hike drastically. Consequently, it seems Turkey's central bank is following other goals than stabilizing the inflation rate through that policy.

Always have been endless arguments among economists about the forecasting power of non-linear volatility models and linear autoregressive ones. In this regard, [7] stated that the exchange rates time series tend to display a long memory behavior; hence, ARFIMA models proposed by [8] should be employed for forecasting purposes. In contrast, [9] argued that the GARCH models, in most cases, beat the linear alternative ones. However, in the case of highly fluctuating data, autoregressive models provide better forecasts. In another extensive empirical study, [10] analyzed Jamaica's exchange rates using a collection of GARCH models. They found out the returns time series had a long memory characteristic and, at the same time, exhibited evidence of asymmetric volatility behavior. Finally, they concluded a GARCH model with a leptokurtic distributed error term could provide the best forecast.

Now, the question is that what model can provide the best description of the fluctuating trends of the Turkish lira exchange rates and would obtain the most accurate forecast? From May until November 2022, the interest rate has deliberately declined two times. This monetary policy, from an econometrics point of view, can cause two structural breaks in the time series overall trend. Therefore, the possible effect of breakpoints should be considered in model tuning. However, in most forecasting studies, the impact of structural breaks has been neglected. The present study has attempted to provide the best possible short-term forecast for the Turkish lira exchange rates by employing univariate models. It should be mentioned this study does not want to claim that the multivariate models are inapplicable in daily return forecasts. Although, the macro-econometric models, by definition, did not design for daily data.

---

[2] www.fred.stlouisfed.org
[3] www.tcmb.gov.tr

On the other hand, there is a growing tendency among econometricians toward employing highly sophisticated models to forecast financial time series. Among all the models, the first rank belongs to artificial neural networks, at least during the past ten years. For instance, [11], [12], and [13] has worked exclusively on the exchange rate forecasts using hybrid neural networks. However, as professor Friedman had stated, the models should be compared through forecasting power, not how much they satisfy assumptions or are complicated [14]. In fact, imposing more assumptions to build a complex model make it unrealistic. Market agents generally do not employ logical and rational trading methods because, in that case, the market should be efficient in the sense of [15] theory. But plenty of studies have shown that financial markets are inefficient. That means most market agents use straightforward calculations, are highly emotional in the sense of herding behavior, and are retrospective with a short decision horizon. As a result, this study has not attempted to forecast the exchange rate through a hybrid complicated model but has tried to test if decreasing the complexity of the model leads to better forecasts and whether the traditional approaches still work well.

The structure of the study is as follows. The second section has provided an initial understanding of time series behavior through basic statistical tests and data transformations. In the third part, several models have been estimated concerning the available sample. In the fourth section, and after the out-of-sample forecast, the models have been compared using three goodness-of-fit criteria. Finally, the concluding remark and some suggestions for further research are presented in the fifth part.

## II) Data and Variable

This study has analyzed the daily USD/TRY exchange rates for six months, from May to December 2022. The sample has been gathered from www.exchangerates.org.uk, which is a reliable database for financial time series. The data set has been divided into two sub-groups of train and the test, with a ratio of 85-15 percent. This investigation has used the daily returns computed through the following formula:

$$Return_t = \frac{Rate_t - Rate_{t-1}}{Rate_{t-1}} * 100 \qquad (1)$$

Generally, financial time series follow a random walk process, with a high level of variance. The mentioned transformation decreases the variance and pushes data toward a homoscedastic condition, making time series more suitable for further analyses by eliminating the possible unit root. Nevertheless, there is a high chance the returns time series contains a unit or a fractional root.

The first step in data analysis is to compute the fundamental statistics like four primary moments and investigate if the data follow a specific distribution. However, before that, it would be beneficial to illustrate the data graph, which is plotted in Figure 1.

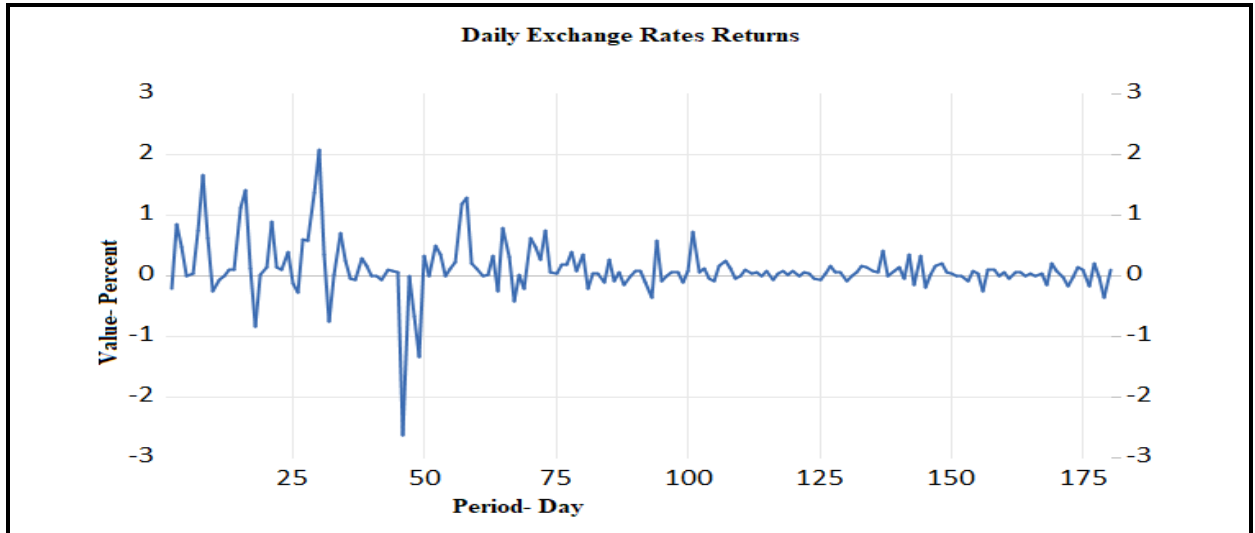

**Figure. 1**. Returns Time Series

It can be deduced from Figure 1 that the time series has a volatile nature. However, the volatilities are more visible in the first third of the chart. Another clear thing is two spikes during the first fifty observations. At first glance, they look like just two outliers, and it is highly predictable that the detection tests confirm it. In fact, if an outlier detection test is utilized, it would find more than just two points, which is unlikely in series with short time intervals. Furthermore, it is obvious the data has a skewed distribution; hence, assigning a normal distribution would not be realistic. Therefore, a high level of fluctuation in the time series trend, especially in the initial section, can be imputed to the non-linear nature of the series. In following, to evaluate the mentioned analyses, descriptive statistics are reported in Table 1.

**Table. 1** *Descriptive Statistics*

| Statistic | Value |
|---|---|
| Mean | 0.109 |
| Median | 0.055 |
| Mode | 0.000 |
| Max | 2.073 |
| Min | -2.645 |
| Std. Dev | 0.436 |
| Skewness | -0.321 |
| Kurtosis | 15.036 |
| J-B Stat. | 1083.451 |
| J-B Prob. | 0.000 |

Source: *Research Findings*

First, it should be mentioned the *J-B* is an abbreviation for the [16] normality test. Second, the *J-B* test shows the data does not follow a normal distribution. Third, as can be deduced from the Table 1, the time series is left-skewed, contrasted to most financial time series that follow right-skewed distributions. On the other hand, skewness is an essential aspect of financial time series because it can be considered a measure of risk [17]. It is crucial because most econometrics models like [18] or ARCH models of [19] assume the data follow normal or symmetric heavy-tailed distributions, which do not comply with this time series condition. However, there are many explanations for negative skewness. For instance, [20] discussed that a stochastic bubble could be the source of left skewness. Therefore, if the time series follows a bubbly regime, using equilibrium models like ARMA or almost all of the pricing theories like the Capital Asset Pricing Model or Arbitrage Pricing Theory, will lose their applicability. As a result, it can be claimed a univariate model cannot describe all the time series fluctuations. To achieve a better understanding, a deep analysis of descriptive statistics will be needed. Since the time series mode is zero, discriminating data based on this criterion would be beneficial. Table 2 is provided an analysis according to the mode cut-off.

**Table. 2** *Frequency Discrimination*

| Value | No. of Days | Percent |
|---|---|---|
| Zero | 16 | 8.94 |
| Negative | 48 | 26.81 |
| Positive | 115 | 64.25 |
| Total | 179 | 100 |
| Max. days in negative returns | | 3 |
| Max. days in positive returns | | 11 |
| Max. days in an increasing trend | | 4 |
| Max. days in a decreasing trend | | 4 |

Source: *Research Findings*

According to Table 2, investors on most days gained positive returns (but not necessarily excess returns). Moreover, the skewness can be easily described since the maximum duration of the period with continuous positive returns was higher than the negative one. Furthermore, the maximum number of days the time series was in an upward movement was equal to the spending days in a downward trend. Thus, it can be concluded that the effect of positive and negative shocks was similar, and hence, there is no leverage effect in the data. It is an important concept because, in the presence of the leverage effect, the data variance tends to behave conditionally, especially if the time series distribution is leptokurtic. To find more evidence, it can be beneficial to calculate the correlation coefficient between $R_t^2$ and $R_{t-1}$ where $R_t$ stands for the return time series. A negative value can be translated to a leverage effect; however, the estimated value is $Corr(R_t^2, R_{t-1}) = 0.1896$. Therefore, it can be said there is no leverage effect.

For further analysis, another test known as the Runs test of [21], which analyzes the data distribution, can be employed. The null hypothesis states the time series follows an identically independent distribution; nevertheless, there is no specification under the alternative hypothesis. The rejection of the null implies the data does not follow a specific distribution but provides some evidence of unit roots as a stochastic time trend. If the data is not covariance stable, the mean and variance are time-dependent, and hence, the data cannot follow an identical independent distribution. It should be mentioned the Runs test is nothing more than a necessary condition for stationarity. As a result, deciding on the stochastic time trend relying only upon the Runs test would lead to false conclusions. The outcomes of the test based on three thresholds are reported in Table 3.

**Table. 3** *Runs test Outcomes*

| Threshold | Mean | Median | Mode |
|---|---|---|---|
| $R$ | 63 | 78 | 86 |
| $\bar{R}$ (Exp.) | 80.776 | 90.497 | 83.235 |
| Std. Dev | 5.942 | 6.670 | 6.126 |
| Z-Stat. | -2.992 | -1.873 | 0.451 |
| Prob. | 0.003 | 0.061 | 0.652 |

Source: *Research Findings*

Since the time series length is more than 20 observations, the critical value has extracted from the standardized normal distribution. As can be seen in Table 3, the null hypothesis has not been rejected for median and mode, while it has been rejected for mean threshold. So, it can be accounted as a sign of a near-stationary process. That means the data does not follow a pure stationary process, but there is a mean-reverting behavior, and hence the effect of a shock decays at a rate slower than a completely stable process [22].

The next step is to determine if the time series contain any time trends. Fortunately, there is a massive statistical literature about those tests; however, this paper has used two of the most well-cited unit root tests, including the augmented Dickey-Fuller (ADF) test of [23] and [24], known as the P-P test. The famous test of [25], known as the KPSS, has been employed to check if the series is stationary. The results of the tests are reported in Table 4.

**Table. 4** *Unit Root/Stationary Tests*

**ADF Test**
Null: There is a Unit Root.
Sig. Level: **5 %**

| Type | Statistic | Critical Value | Prob. |
|---|---|---|---|
| Pure | -9.565 | -1.943 | 0.000 |
| Intercept and Trend | -10.356 | -3.435 | 0.000 |

**P-P Test**
Null: There is a Unit Root.
Sig. Level: **5 %**

| Type | Statistic | Critical Value | Prob. |
|---|---|---|---|
| Pure | -9.537 | -1.943 | 0.000 |
| Intercept and Trend | -10.320 | -3.435 | 0.000 |

**KPSS Test**
Null: Time Series is Stationary
Sig. Level: **5 %**

| Type | *L-M* Stat. | Critical Value |
|---|---|---|
| Intercept | 0.513 | 0.463 |
| Intercept and Trend | 0.159 | 0.146 |

Source: *Research Findings*

The outcomes of Table 4 have asserted the time series neither has a unit root nor is stationary. This behavior could be a sign of a long-memory trend in time series, which is in the direction of the Runs test outcomes. However, for modeling purposes, stationarity of the time series is an essential condition. Thus, to obtain a stable series, the data should be subject to a first-order differencing. The KPSS test results on the new time series are reported in Table 5.

**Table. 5** *Stationary Test*

**KPSS Test**
Null: *D* (Return) Time Series is Stationary
Sig. Level: **5 %**

| Type | *L-M* Stat. | Critical Value |
|---|---|---|
| Intercept | 0.095 | 0.463 |
| Intercept and Trend | 0.087 | 0.146 |

Source: *Research Findings*

As can be deduced from Table 5, the series has got stationary after differencing. As a neglected point, if there are structural breaks in the time series, the diagnostic power of stability tests will be decreased exponentially. The first time, [26] showed while the series has a unit root, the test rejected the null hypothesis in favor of the alternative. For this purpose, he added a dummy variable to the restricted model of the ADF test and did the test. He concluded that structural breaks increase type I error in unit root tests. In this regard, it is vital to examine the time series for possible structural breaks. However, well-cited tests like [27] assume the breakpoint is known, which is not our case. To overcome that issue, [28, 29] suggested a test that does not need any prior information about the breakpoints. The test results are reported in Table 6.

**Table. 6** *Structural Breaks Test*

**Multiple Breakpoints Tests**
**Bai-Perron** Test of *L vs. L+1* Sequentially Determined Breaks

Breaking Variable: *C* (*Level*)

Max Breaks: **5**    Sig. Level: **0.05**

Covariance: Heteroscedastic-Autocorrelation Consistent (**HAC**)

| Break Test | *F*-Stat. | Scaled *F*-Stat. | Critical Value |
|---|---|---|---|
| 0 *vs.* 1 | 3.563 | 3.563 | 8.580 |

Source: *Research Findings*

The above results show the test statistic is less than the critical value; thus, the null of no breakpoint cannot be rejected. As a result, since there is no structural break, the outcomes of unit root/stability tests are reliable.

### III) Modeling and Estimation

In almost all the univariate analyses, the autoregressive-moving average model, ARMA, is the first-line model for forecasting purposes. The model assumes the underlying process is linear and stationary. The general specification of an ARMA(p, q) model is as follows:

$$\varphi(L)Y_t = \theta(L)\varepsilon_t \quad s.t. \quad \varepsilon_t \sim WN(0, \sigma^2) \quad (2)$$

Where *L* is the lag operator in which $L(Y_t) = Y_{t-1}$ and $\varepsilon_t$ are errors known as moving average components. Both $\varphi(L)$ and $\theta(L)$ are two polynomials of *L* from the orders of *p* and *q*, respectively. If the absolute value of all the roots of the $\varphi(L)$ is greater than one, then the model is stationary. In other cases, the time series needs to get stable through the differentiation operator as the following process:

$$\omega(L)(1-L)^d Y_t = \theta(L)\varepsilon_t \quad s.t. \quad (1-L)Y_t = \Delta Y_t = Y_t - Y_{t-1} \quad (3)$$

And the parameter *d* is the number of differentiations needed to obtain a stationary process. In this case, the model is called ARIMA(p, d, q).

The starting step in constructing an ARMA model is to decide on the number of lags. In this regard, taking advantage of the autocorrelation (ACF) and partial autocorrelation (PACF) functions would be beneficial. The graph of the functions is illustrated in Figure 2.

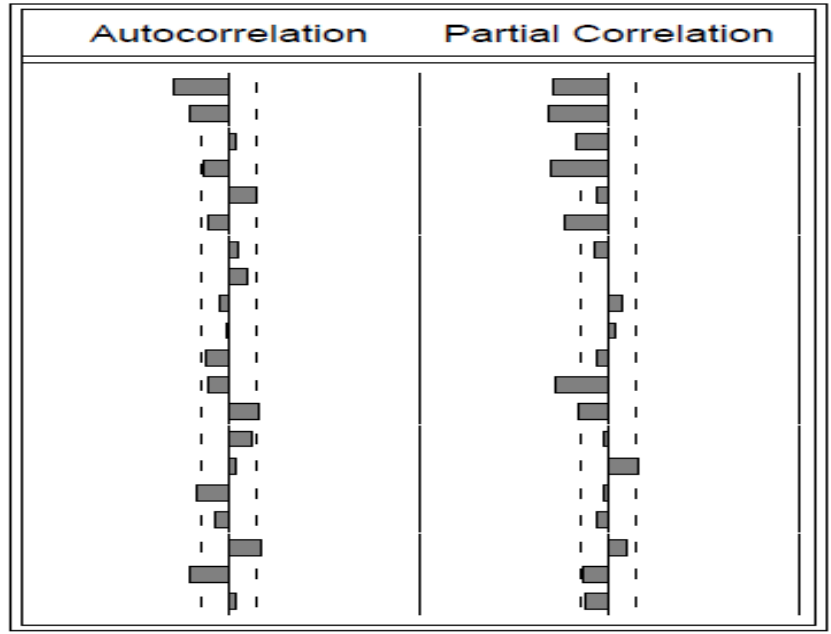

**Figure. 2**. ACF and PACF of Differentiated Time Series
**Source:** *Research Findings*

As can be seen, there are two significant spikes in the ACF part, so it suggests an MA(2) process. In contrast, the PACF indicates significant jumps in lags one, two, four, and six. That means there are three possibilities for the autoregressive part, including AR(2), AR(4), and AR(6). However, for a better decision on the autoregressive component, taking advantage of information criteria would be helpful. The selected models concerning the criteria are reported in Table 7.

**Table. 7** *Model Selection*

| Model: ***ARMA (P, Q)*** $s.t.$ $P, Q \in [1, 7]$ | | |
|---|---|---|
| Variable: **D (Returns)** | | |
| Estimating Technique: Maximum Likelihood **(Normal Distribution)** | | |
| Criterion | Value | Suggestion |
| *AIC* | 1.097 | ARMA(7, 2) |
| *BIC* | 1.205 | ARMA(2, 2) |
| *H-Q* | 1.159 | ARMA(1, 2) |

Source: *Research Findings*

According to Table 7, all the criteria indicate the moving average part should contain two lags, as has been confirmed by the ACF plot. However, there is no unique agreement about the autoregressive element. On the other hand, among the criteria, only the BIC has offered lags similar to the PACF function. It was not a surprising outcome because, as a general rule, the Bayesian criterion is more suitable for small samples [30]. As a result, two lags have been picked for the autoregressive component. The estimated model is reported in Table 8.

**Table. 8** *Estimated ARIMA Model*

Estimated Model: *ARIMA(2, 1, 2)*
Estimating Method: **Maximum Likelihood (Normal Distribution)**

| Variable | Coef. | Std. Err. | t-Stat. | Prob. |
|---|---|---|---|---|
| C | -0.002 | 0.002 | -0.913 | 0.362 |
| AR(1) | -0.534 | 0.140 | -3.807 | 0.000 |
| AR(2) | 0.092 | 0.083 | 1.101 | 0.272 |
| MA(1) | -0.152 | 141.724 | -0.001 | 0.999 |
| MA(2) | -0.848 | 1030.836 | -0.001 | 0.999 |
| *R-Sq* | 0.401 | | | |
| Adj. *R-Sq* | 0.384 | | AIC | 1.128 |
| *F*-Stat. | 23.079 | | BIC | 1.205 |
| *F*-Prob. | 0.000 | | H-Q | 1.171 |

**Normality Test**

| | | | | |
|---|---|---|---|---|
| *J-B* Stat. | 1397.033 | | | |
| *J-B* Prob. | 0.000 | | | |

**Heteroscedasticity Test: ARCH**

| | | | | |
|---|---|---|---|---|
| *F*-Stat. | 1.335 | Prob. *F*(1, 175) | 0.249 | |
| *Lagrange*-Stat. | 1.340 | Prob. *Chi-Sq*(1) | 0.247 | |

Source: *Research Findings*

Unfortunately, the estimation has come with disappointing results both in explanatory power and residual diagnosis. First, except for the AR(1), all the coefficients were statistically insignificant. Moreover, the adjusted R-squared statistic is around 39 percent. The worst part was residual diagnostic test outcomes because it displayed error terms have not been distributed normally. However, the ARCH effect has been rejected, and thus, the residuals are homoscedastic. Also, another part that has not been reported in Table 8 is the serial correlation test. For this purpose, the Q-statistic of [31] for lags from five to ten has been analyzed. Since the estimated parameters ($P + Q = 4$) have restricted the degrees of freedom, the test has to start from the fifth lag. The outcomes are reported in Table 9.

**Table. 9** *Ljung-Box test*

| Variable: *ARIMA(2, 1, 2)* Residuals | | |
|---|---|---|
| **Null**: There is no Serial Correlation. | | |
| Sig. Level: **0.05** | | |
| Lag | Q-Stat. | Prob. |
| 5 | 2.245 | 0.134 |
| 6 | 2.248 | 0.325 |
| 7 | 3.078 | 0.380 |
| 8 | 4.728 | 0.316 |
| 9 | 5.578 | 0.350 |
| 10 | 7.647 | 0.265 |

Source: *Research Findings*

The results reported in Table 9 show there is no autocorrelation in the residuals' time series. As stated earlier, the tests in Table 8 rejected the ARCH effect as a kind of heteroscedasticity. As a result, there is no reason to estimate a GARCH model. The last family of models that will be discussed is known as the exponential smoothing models, which were first introduced by [32]. The models (Known as Brown's models) assume the closer data have more influence on the overall trend than the distant data. Suppose $\{Y_t\}_1^N$ is a time series and the exponentially smooth filtered values are denoted by $\hat{Y}_t$. Also, consider an adjusting parameter $0 \leq \alpha \leq 1$ where it controls the weighting procedure. Moreover, suppose there is no deterministic time trend and seasonality. Therefore, the filtering process is as follows:

$$\hat{Y}_t = \alpha Y_t + (1-\alpha)\hat{Y}_{t-1} \quad s.t. \quad \hat{Y}_1 = Y_1 \quad (4)$$

Hence, for the forecasting purposes, $Y_{t+h|t} = \hat{Y}_t$.

[33] extended Brown's model by adding a linear trend in the time series (known as Holt's model), and rewrite the equations as follows by introducing a new parameter $\beta \in [0, 1]$.

$$\hat{Y}_t = \alpha Y_t + (1-\alpha)(\hat{Y}_{t-1} + \Delta\hat{Y}_{t-1}) \quad s.t. \quad \hat{Y}_1 = Y_1$$

$$\Delta\hat{Y}_t = \beta(\hat{Y}_t - \hat{Y}_{t-1}) + (1-\beta)\Delta\hat{Y}_{t-1} \quad s.t. \quad \Delta\hat{Y}_2 = Y_2 - Y_1 \quad (5)$$

Where similar to Brown's model, $\alpha$ is the smoothing parameter for the level factor and $\beta$ is the smoothing parameter for the trend. Forecasting in this framework is quite simple using the below equation:

$$Y_{t+h|t} = \hat{Y}_t + h.\Delta\hat{Y}_t \quad (6)$$

It should be mentioned if $\beta = 0$, Holt's model reduces to Brown's specification. There are two approaches to determining control parameters, including assigning values to the data prior to the calculation or using goodness-of-fit criteria like root mean squared error (RMSE) or mean absolute error (MAE). For instance, [34] suggested an $\alpha \in [0.1, 0.3]$ would be a suitable choice, however, [35] discussed in favor of a parameter on the interval of $[0.05, 0.5]$. In contrast, [36] stated that using forecasting evaluation criteria could provide better model tuning compared to assigning prior beliefs on the parameters. Nevertheless, In order to avoid any possible bias, this study has used the RMSE criterion. The estimated elements of Holt's and Brown's models are reported in Table 10.

**Table. 10** *Exponential Smoothing Estimation*

| Variable: **Returns** | | | | |
|---|---|---|---|---|
| No. of Observations: **179** | | | | |
| Model | Alpha (α) | Beta (β) | Sum Sq-Resid. | RMSE |
| Holt | 0.160 | 0.000 | 35.475 | 0.445 |
| Brown | 0.026 | -- | 33.619 | 0.433 |

Source: *Research Findings*

As can be seen, the beta parameter in Holt's model is zero, and hence, there is no deterministic trend. Therefore, the method is equivalent to Brown's technique. Accordingly, only Brown's exponential smoothing method will be employed. The interesting point about the estimated parameters is that although the beta is zero, and so, by definition, two models should be equivalent, the estimated alphas are different. The reason behind this disagreement is that a beta equal to zero reduces the value of $\Delta \hat{Y}_{t-1}$ to $\Delta \hat{Y}_2 = Y_2 - Y_1$, but in the time series under investigation, $Y_2 \neq Y_1$; therefore, the two models are not completely equal. However, since the beta is zero, the estimated alpha for Holt's model should be neglected.

There are several concerns regarding the estimated models. First, none of the estimated models have provided normally-distributed residuals. This phenomenon can be related to some omitted variables or the model misspecification. Second, all models' explanatory power (adjusted R-square statistic) was less than fifty percent. Consequently, the estimated models have no ability to explain all the underlying reasons behind the fluctuating behavior of the time series. In summary, the estimated models of this study should only be applied for short-term forecasting purposes.

**IV) Forecasting and Discussion**

As discussed earlier, the ARIMA(2, 1, 2) has been selected as the most suited model. However, as suggested by the PACF plot, two other lags, including four and six, also could be picked as the autoregressive part. Hence, ARIMA(4, 1, 2) and ARIMA(6, 1, 2) also will be used in the forecasting stage. To widen this domain, four other models, including MA(2), AR(2), AR(4), and AR(6), have been added to the models' collection. Furthermore, as a tradition in financial time series forecast, the random walk model, which is the symbol of the efficient market hypothesis (EMH), likewise has been employed in order to be the benchmark model. While this hypothesis has been subjected to several criticisms[4], it is still considered a suitable model for comparison purposes. In fact, forecasting through this model is quite simple, and for this reason, it is called the naïve forecasting procedure. In this algorithm, all the approximated future data are equal to the last observation. Mathematically speaking,

$$\hat{Y}_{t+h|t} = Y_t \quad where \quad h = 1,2,... \quad (7)$$

---

[4] For instance, [37], by estimating the Hurst exponent in a rolling-windows procedure, rejected a random walk hypothesis in favor of a fractal one in the Warsaw stock exchange.

The equation holds because:

$$Y_t = Y_{t-1} + \varepsilon_t \quad s.t. \quad \varepsilon_t \sim WN(0, \sigma^2)$$

$$\rightarrow E(Y_t | Y_{t-1}) = E(Y_{t-1} | Y_{t-1}) + E(\varepsilon_t) \rightarrow E(Y_t | Y_{t-1}) = Y_{t-1} \quad (8)$$

Another simple forecasting model is the mean indicator. The model is almost similar to the naïve forecast, but the last observation should be replaced with the sample mean. Roughly speaking, all the future values equal the time series expected value. Thus,

$$\hat{Y}_{t+h|t} = E(Y_t) \quad where \quad h = 1, 2, \dots \quad (9)$$

The last model is exponential smoothing, and according to the previous section, only Brown's model will be used.

This study has employed three evaluating criteria, including RMSE, MAE, and the symmetric Mean Absolute Percentage Error (SMAPE). All three indices calculate as follows:

$$RMSE = \sqrt{\frac{\sum_{i=1}^{n}(\hat{Y}_i - Y_i)^2}{n}} \quad (10)$$

$$SMAPE = \frac{\sum_{i=1}^{n} \frac{|\hat{Y}_i - Y_i|}{|\hat{Y}_i| + |Y_i|}}{n} * 200 \quad (11)$$

$$MAE = \frac{\sum_{i=1}^{n} |\hat{Y}_i - Y_i|}{n} \quad (12)$$

By definition, all the criteria are non-negative, and a value equal to zero means a fully matched forecast. As a result, values closer to zero indicate better forecasting performance. However, since RMSE uses a square operator, it is more sensitive to outliers compared to MAE. On the other hand, if both the actual value and the forecasted one be too close to zero, the symmetric MAPE could be undefined. Thus the MAE is the study's preferred criterion.

The test group for the out-of-sample forecast contained 33 observations, and the models forecasting evaluation is reported in Table 11.

**Table. 11** *Forecasting Evaluation*

| Model | RMSE | MAE | SMAPE |
|---|---|---|---|
| ARIMA(2,1,2) | 0.218 | 0.154 | 158.847 |
| ARIMA(4,1,2) | 0.221 | 0.156 | 159.427 |
| ARIMA(6,1,2) | 0.228 | 0.156 | 154.586 |
| AR(2) | 0.209 | 0.147 | 156.286 |
| AR(4) | 0.221 | 0.155 | 159.103 |
| AR(6) | 0.231 | 0.158 | 162.342 |
| MA(2) | 0.218 | 0.153 | 157.511 |
| Random Walk | 0.217 | 0.144 | 140.263* |
| Mean Index | 0.226 | 0.157 | 141.562 |
| Brown's Smoothing | 0.205* | 0.123* | 147.414 |

The * indicates the best model.
Source: *Research Findings*

Table 11 clearly shows no agreement among criteria in selecting the best forecasting model. However, Two out of three criteria have selected the simple exponentially weighted method. It should be mentioned the naïve forecast is the second-best model, and this finding confirms that a random walk model should always be among the forecasting techniques.

The outcomes asserted the ARIMA models exhibited poor performance in the forecasting step. But, it could be attributed to the fact that the time series has displayed some evidence of long memory behavior, which complies with [7] conclusion. On the other hand, the lack of heteroscedasticity contradicts the findings of [10]. Finally, the weakness of autoregressive models in the prediction stage is contrary to the argument of [9].

## V) Conclusion

Forecasting financial time series has always been a desired task for all market agents, especially risk-taker investors. Forecasting makes investors capable of seeing beyond the uncertainty surrounding future trends and thus taking advantage of numerous opportunities. Among all the financial time series, the exchange rates have a unique place because it is not only considered a vital parameter in monetary policies but highly correlated with the citizens' daily life. A sharp decline in national currency decreases purchasing power by increasing the prices of imported goods and services. Moreover, its inflationary effects does not restrict to imported commodities because it causes foreign trade imbalance and eventually pushes living standards toward lower levels.

Turkey always has a unique position among its neighbors due to its situation as a bi-continent country and the fact that it is one of the safest transit channels between Europe and Western Asia. As a result, the country plays a crucial role in regional and international economics. Nevertheless, during the recent past year, the Turkish lira exchange rate has been subjected to severe fluctuations and caused several inflationary waves. In fact, multiple reasons, including the increase in the world's inflation levels due to Russia's invasion of Ukraine or economic recession of the Covid-19 pandemic, can be considered a trigger for such a variable trend. However, among all the nominated explanations, the Turkish central bank policy to decline interest rate has the most influence. However, As much as volatility in exchange rates could be harmful to macroeconomics, it provides a golden situation for some investors to take advantage of the arbitrage opportunities. The subject got more interesting when, in the absence of inflation-indexed government bonds, new amateur investors entered into the market with the desire to hedge their savings against the upcoming inflationary waves. In this regard, the present study has employed several univariate models to provide reliable forecasting using the USD/TRY daily time series. For this purpose, the linear models of Box-Jenkins and exponentially weighted smoothing techniques have been utilized. Although the time series exhibited left-skewed leptokurtic distribution, the presence of the ARCH and leverage effects have been rejected.

In the modeling step, seven ARIMA models and two types of smoothing filters have been estimated. Since the trend parameter in Holt's method was zero; thus, only Brown's smoothing filter with an alpha near zero has used. Furthermore, to provide a benchmark among the competitive models, this study took advantage of a pure random walk model, well-known as a naïve forecasting procedure. The forecasting evaluation revealed that Brown's method provided the best predictions; however, the second best place was allocated to the naive model.

As a result, the study concluded that the simple models can outperform the sophisticated ones and the traditional forecasting models still have some levels of applicability. In this regard, it suggests that amateur investors in the exchange market should at least use long-established techniques like a random walk model.

During the modeling process, the unit root and stationary tests showed the time series has some characteristics of a long-memory process. Future studies should focus on this feature and determine if the behavior is a true mean-reverting process or just got mistaken with a more complex model of Markov regime switching.

**Declarations**

- The data that support the study findings are available freely and publicly at https://www.exchangerates.org.uk/USD-TRY-exchange-rate-history.html

- The author declares he has no known competing financial interests or personal relationships that could have appeared to influence the work reported in this paper.

- This research did not receive any financial aid from government, private, or not-for-profit agencies.

**References**

1. Fetai, B., Koku, p. S., Caushi, A., Fetai, A. The Relationship between Exchange Rate and Inflation: The Case of Western Balkans Countries. *Journal of Business Economics and Finance*. 2016; 5 (4): 360-364.

2. Edison, H. J. Purchasing Power Parity in the Long run: A Test of the Dollar/Pound Exchange Rate (1890-1978). *Journal of Money, Credit and Banking*. 1987; 19 (3): 376-387.

3. Thahara, A. F., Rinosha, K. F., Shifaniya, A. J. F. The Relationship between Exchange Rate and Trade Balance: Empirical Evidence from Sri Lanka. *Journal of Asian Finance, Economics and Business*. 2021; 8 (5): 37-41.

4. Najwa, R., Mansur, M. The Relationship between Exchange Rate and Trade Balance: Evidence from Malaysia based on ARDL and Nonlinear ARDL, University Library of Munich MPRA Paper. 2018; (112447).

5. Kouladoum, J. – C. External Debts and Real Exchange Rates in Developing Countries: Evidence from Chad. University Library of Munich MPRA Paper. 2018; (88440).

6. Christensen, I., Dion, F., Reid, C. Real Return Bonds, Inflation Expectations, and the Break-Even Inflation Rate. Bank of Canada Working Paper. 2004; (2004-43).

7. Cheung, Y. –W. Long Memory in Foreign-Exchange Rates. *Journal of Business & Economic Statistics*. 1993; 11 (1): 93-101.

8. Granger, C. W. J., Joyeux, R. An Introduction to Long-Memory Time Series Models and Fractional Differencing. *Journal of Time Series Analysis*. 1980; 1 (1): 15-29.

9. Bollerslev, T., Wright, J. H. High Frequency Data, Frequency Domain Inference, and Volatility Forecasting. *The Review of Economics and Statistics*. 2001; 83 (4): 590-602.

10. Longmore, R., & Robinson, W. Modeling and Forecasting Exchange Rate: An Application of Asymmetric Volatility Models. Bank of Jamaica Working Paper. 2004; (WP 2004/03).

11. Pacelli, V., Bevilacqua, V., Azzollini, M. An Artificial Neural Network Model to Forecast Exchange Rates. *Journal of Intelligent Learning Systems and Applications*. 2011; 3 (2): 57-69.

12. Khashei, M., Torbat, S., Haji Rahimi, Z. An Enhanced Neural-based Bi-Component Hybrid Model for Foreign Exchange Rate Forecasting. *Turkish Journal of Forecasting*. 2017; 1 (1): 16-29.

13. Pfahler, J. F. Exchange Rate Forecasting with Advanced Machine Learning Methods. *Journal of Risk and Financial Management*. 2022; 15 (2): 15010002.

14. Friedman, M. The Methodology of Positive Economics. In: Essays in Positive Economics. Chicago: University of Chicago Press; 1953. 3-43 p.

15. Fama, E. F. A Review of Theory and Empirical Work. *The Journal of Finance*. 1970; 25 (2): 383-417.

16. Jarque, C. M., Bera, A. K. Efficient Tests for Normality, Homoscedasticity and Serial Independent of Regression Residuals. *Economic Letters*. 1980; 6 (3): 255-259.

17. Grigoletto, M., Lisi, F. Looking for Skewness in Financial time Series. *The Econometrics Journal*. 2009; 12 (2): 310-323.

18. Box, G., Jenkins, G. Time Series Analysis: Forecasting and Control. San Francisco: Holden-Day; 1970.

19. Engle, R. E. Autoregressive Conditional Heteroscedasticity with Estimates of the Variance of United Kingdom Inflation. *Econometrica*. 1982; 50 (4): 987-1007.

20. Blanchard, G. J., Watson, M. W. Bubbles, Rational Expectations and Financial Market. National Bureau of Economic Research (NBER) Working Paper. 1982; (945).

21. Wald, A., Wolfowitz, J. On a Test Whether Two Samples are from the same Population. *Annals of Mathematical Statistics*. 1940; 11 (2): 147-162.

22. Baillie, R. T. Long Memory Processes and Fractional Integration in Econometrics. *Journal of Econometrics*. 1996; 73 (1): 5-59.

23. Said, E. S., Dickey, D. A. Testing for Unit Roots in Autoregressive Moving Average Models of Unknown Order. *Biometrica*. 1984; 71 (3): 599-607.

24. Phillips, P. C. B., Perron, P. Testing for a Unit Root in Time Series Regression. *Biometrika*. 1988; 75 (2): 335-346.

25. Kwiatkowski, D., Phillips, P. C. B., Schmidt, P., Shin, Y. Testing the Null Hypothesis of Stationary against the Alternative of a Unit Root: How sure are we that Economic Time Series have a Unit Root? *Journal of Econometrics*. 1992; 54 (1-3): 159-178.

26. Perron, P. The Great Crash, the Oil Price Shocks, and the Unit Root Hypothesis. *Econometrica*. 1989; 57 (6): 1361-1401.

27. Chow, G. C. Tests of Equality between Sets of Coefficients in two Linear Regressions. *Econometrica*. 1960; 28 (3): 591-605.

28. Bai, J., Perron, P. Estimating and Testing Linear Models with Multiple Structural Changes. *Econometrica*. 1998; 66 (1): 47-78.

29. Bai, J., Perron, P. Computation and Analysis of Multiple Structural Change Models. *Applied Econometrics*. 2003; 18 (1): 1-22.

30. Enders, W. Applied Econometric Time Series (Fourth Edition). New York: Wiley; 2014. 76 p.

31. Ljung, MG. M., Box, G. E. P. (1978). On a Measure of Lack of Fit in Time Series Models. *Biometrika*. 1978; 65 (2): 297-303.

32. Brown, R. G. Exponential Smoothing for Predicting Demand. Massachusetts: Arthur D. Little Inc.; 1956. 15 p.

33. Holt, C. C. Forecasting Trends and Seasonality by Exponentially Weighted Average. *International Journal of Forecasting*. 1957; 20 (1): 5-10.

34. Jacobs, F. R., Chase, R. B. (2013). Operation and Supply Chain Management: The Core (Third Edition), New York: McGraw-Hill Higher Education; 2013. 59 p.

35. Heizer, J., & Render, B. Operation Management (Tenth Edition). New Jersey: Prentice Hall; 2011. 127 p.

36. Chopra, S., Meindl, P. Supply Chain Management: Strategy, Planning and Operation (Fifth Edition). New Jersey: Prentice Hall; 2013. 195 p.

37. Raeisi Sarkandiz, M., Bahlouli, R. The Stock Market between Classical and Behavioral Hypotheses: An Empirical Investigation of the Warsaw Stock Exchange. *Econometric Research in Finance*. 2019; 4 (2): 67-88.